\documentclass[useAMS,usenatbib]{mn2e}
\usepackage{graphicx}

\def\deg{\ifmmode^\circ\else$^\circ$\fi}
\def\kms{km\thinspace s$^{-1}$}
\def\msun{M$_{\odot}$}
\def\lsun{L$_{\odot}$}

\newcommand{\Lsolar}{L$_{\odot}$}
\newcommand{\mic}{\,$\mu$m}
\newcommand{\dg}{$^{\circ}$}

\title[ WFCAM, Spitzer-IRAC and SCUBA photometry of DR21/W75]
{WFCAM, Spitzer-IRAC and SCUBA observations of the massive star forming
region DR21/W75: II. Stellar content and star formation}
\author[M.S.N.~Kumar et al.]
       {M.S.N.~Kumar$^{1}$, 
	C.J.~Davis$^{2}$, 
	J. M. C.~Grave$^{1,5}$,
	B.~Ferreira$^{3}$ \& 
	D.~Froebrich$^{4}$\\
$^1$Centro de Astrofísica da Universidade do Porto,
    Rua das Estrelas s/n 4150-762 Porto, Portugal \\
$^2$Joint Astronomy Centre, 660 North A'oh\={o}k\={u} Place,
        University Park, Hilo, Hawaii 96720, USA. \\
$^3$Department of Astronomy, University of Florida, Gainesville, FL 32611-2055, USA\\
$^4$Dublin Institute for Advanced Studies, 5 Merrion Square, Dublin 2, Ireland \\
$^5$Departamento de Matematica Aplicada da Faculdade de Ci\^{e}ncias da Universidade do Porto, Portugal\\
}
\begin{document}

\date{ }

\pagerange{\pageref{firstpage}--\pageref{lastpage}} \pubyear{2006}

\maketitle

\label{firstpage}

\begin{abstract} Wide field near-infrared observations and Spitzer
Space Telescope IRAC observations of the DR21/W75 star formation
regions are presented. The photometric data are used to analyse the
extinction, stellar content and clustering in the entire region by
using standard methods. A young stellar population is identified all
over the observed field, which is found to be distributed in embedded
clusters that are surrounded by a distributed halo population
extending over a larger projected area. The Spitzer/IRAC data are used
to compute a spectral index value, $\alpha$, for each YSO in the
field.  We use these data to separate pure photospheres from disk
excess sources. We find a small fraction of sources with $\alpha$ in
excess of 2 to 3 (plus a handful with $\alpha\sim$4), which is much
higher than the values found in the low mass star forming region IC348
($\alpha \le 2$).  The sources with high values of $\alpha$ spatially
coincide with the densest regions of the filaments and also with the
sites of massive star formation. Star formation is found to be
occuring in long filaments stretching to few parsecs that are
fragmented over a scale of $\sim$1\,pc. The spatial distribution of
young stars are found to be correlated with the filamentary nebulae
that are prominently revealed by 8$\mu$m and 850\mic\ observations.
Five filaments are identified that appear to converge on a center that
includes the DR21/DR21(OH) regions. The morphological pattern of
filaments and clustering compare well with numerical simulations of
star cluster formation by \citet{bate03}.

\end{abstract}

\begin{keywords}
 stars:formation -- ISM:HII regions -- infrared: stars -- turbulence
\end{keywords}

\section{Introduction}

Wide-field near-infrared (NIR) and mid-infrared (MIR) surveys
are potentially very useful for mapping the distribution of young
stars in Giant Molecular Clouds (GMCs).  In particular, the launch of
the Spitzer Space Telescope has brought about a revolution in such
studies, and several nearby, low-mass star forming regions have been
analysed recently \citep[ex][]{jorgenson06,harvey06}.  With the advent
of wide-field imagers on mid-sized, ground based telescopes like the
U.K.  Infrared Telescope (UKIRT), the deeper near infrared data can be
effectively combined with the Spitzer Space telescope observations and
the techniques employed on nearby low mass cores can then be applied
to relatively distant (d$\sim$2--3~kpc), high-mass star forming
regions.

Two such regions are DR21 and W75N in the Cygnus X HII complex.
Traditionally observed separately, DR21 and W75N are associated with
massive, dense cores separated by about half a degree on the sky
\citep{wil90,shi03}. Both sources are associated with
ultra-compact HII regions \citep{has81,cyg03}, water, OH and/or
methanol masers \citep{pla90,hun94,tor97,kog98}, and extremely massive
bipolar outflows \citep{moo91,gar91,dav96,she03}.  DR21(OH), a
third site of massive star formation, lies $\sim$3\arcmin\ north of
DR21.  Although it too is associated with intense maser activity
\citep{kog98,man92}, unlike DR21, it is not detected at
radio   wavelengths  and its  bipolar outflow  is   not  seen in H$_2$
emission.   All  these   regions are  bright  at  Far-Infrared   (FIR)
wavelengths; DR21   is  estimated to   have   an  FIR   luminosity  of
1.5$\times$10$^{5}$\lsun\ and the W75 region 5-8$\times$10$^4$\lsun\
\citep{har77,cam82}. DR21(OH) probably harbours  multiple sources at a
younger stage of evolution.  DR21 and DR21(OH)  coincide with just two
peaks in a  chain  of molecular  cores, traced  in molecular  line and
submillimeter continuum emission,  that  extends north-south over  at
least    12\arcmin\ \citep{cha93,val06}.   \citet{keto90}   found
collapse signatures   in two of   the  main protostellar condensations
coinciding with the UCHII regions DR21 and DR21~D and estimated masses
of  270\msun\ and 20\msun\ respectively. All of the above suggest that
DR21/W75N is a region with intense massive star formation activity.

The DR21/W75 region has recently been observed with Spitzer and the
wide-field camera, WFCAM, at UKIRT.   An overview of the Spitzer IRAC and
MIPS data has been given by \citet{mar04}; \citet{smi05} and
\citet{per06} examine the MIR photometry of the massive, embedded
young stars associated with just the DR21 and W75N cores,
respectively.  The WFCAM observations have been presented by
\citet[][hereafter Paper I]{dav06}, where narrow-band images in
H$_2$ 1-0S(1) emission have been used to search for outflows and jets
across a 0.8\deg $\times$0.8\deg\ field.  In Paper I the Spitzer/IRAC
photometry is used to search for candidate outflow sources; an
850\mic\ mosaic is also presented, which traces the W75N cloud and the
chain of very young (possibly pre-stellar) star-forming cores running
through DR21 and DR21(OH).  A colour image showing the field
observed with Spitzer is shown here in Fig.~1, along with contours of
the 850\mic\ dust continuum emission, for reference. The relationship
between the many outflows, the embedded young stars, the UCHII
regions, masers and molecular cores is discussed in some detail in
Paper I.  In this paper -- Paper II -- we step back and take a more
global view of DR21/W75: the Spitzer/IRAC photometry and UKIRT/WFCAM
JHK data are used to characterize and map the distribution of young
stars across the field.  We derive extinction maps, surface-density
maps, and plot colour-colour and colour-magnitude diagrams for some
43,000 WFCAM sources and 1580 Spitzer targets.  We use these results
to study the nature of the clustered and distributed populations, and
their relationship to dense cores, filaments, and the locations of
massive stars.

\begin{figure*}
\vskip -2cm
\vbox to270mm{\vfil
\includegraphics[width=160mm]{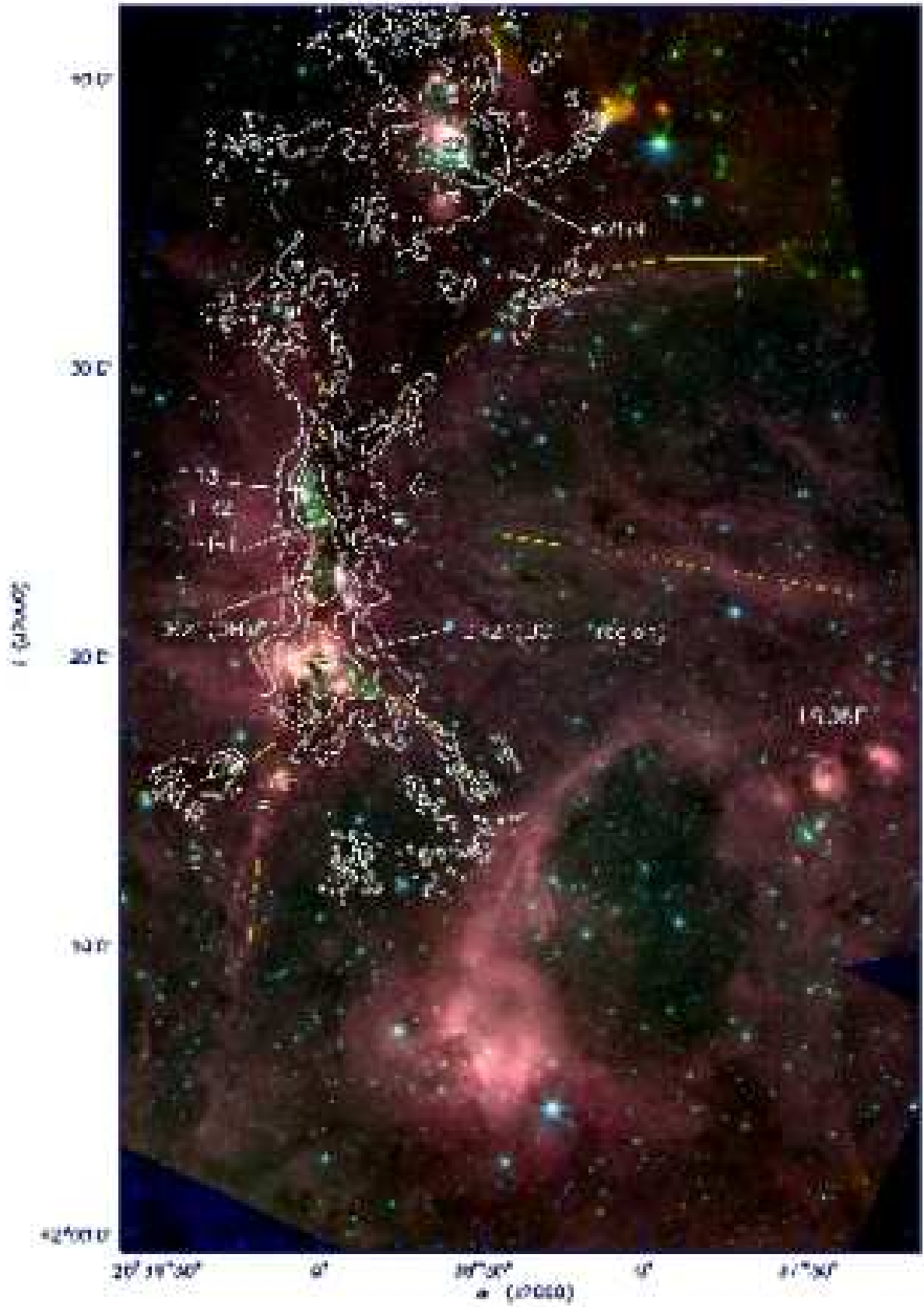}
 \caption{A colour image of the full DR21/W75 mosaic observed with the
 Spitzer Space Telescope, composed from images in 3.6\mic (blue),
 4.5\mic (green) and 8.0\mic (red). Contours represent the 850\mic\
 emission mapped using SCUBA on JCMT. The sketched lines are meant to
 indicate the large scale filamentary emission.}  \vfil}
\label{jhk}
\end{figure*}

DR21 and W75N are associated with clouds with average Local Standard
of Rest (LSR) velocities of $\sim$-3\kms\ and $\sim$+9\kms ,
respectively; a small amount of foreground gas at $V_{\rm
LSR}\sim+9$\kms\ may also lie in front of DR21.  \citet{dic78}
propose that the DR21 and W75N cores are interacting, based on their
modest-resolution mapping and an observed velocity gradient across
W75.  However, in our higher-resolution 850\mic\ maps in Paper I we
fail to detect dust continuum emission between W75N and the 12\arcmin
-long DR21 filament.  Moreover, a distance of $\sim2$~kpc is usually
adopted for W75N, while 2--3~kpc is assumed for DR21 and DR21(OH)
\citep{gen77,cam82,fis85,ode93}.  The higher value gives a more realistic
population of massive stars in our analysis below, so we adopt 3~kpc
as a general distance to DR21, DR21(OH) and W75N regions in this paper.

\section{Data selection and Analysis}

\subsection{JHK Data Selection} 

WFCAM observations of a 0.8\dg $\times$0.8\dg\ field, known as a WFCAM
``tile'', were obtained in Service Observing  time at UKIRT in May and
June of 2005. Data through {\em Mauna Kea  Consortium} broad-band J, H
and  K filters, and  through a narrow-band  H$_2$  1-0S(1) filter were
secured; the H$_2$ data are presented in Paper I, where the details of
the  WFCAM observations  and data  reduction  are also  discussed. The
WFCAM field encompasses  all  of the  region  covered by the Spitzer data
displayed in Fig.~1.

Catalogues of WFCAM point source photometry in the J, H and K bands
were provided by the WFCAM Science archive maintained by the Cambridge
Astronomical Survey Unit (CASU). JHK data were retrieved from the
merged JHK catalog. The following four constraints were used when
querying the catalog: a) magnitude errors of less than 0.2~mag in each
band, b) a merged class identified as a star in the catalog, c) a
probability that the point is a star, given by the {\em pStar}
parameter, of greater than 0.999, and d) coordinates matched to better
than 0.4$\arcsec$ between the JHK bands ($\chi$ and $\eta$ within
$\pm$0.2$\arcsec$ in each band). Regions of star formation contain
large numbers of embedded sources which may not be visible in the J
band.  Therefore, the unmerged catalogs were used to obtain data in
the H and K bands separately to find matching sources only in these
two bands.

The JHK merged catalog with the above constraints yielded 42792 point
sources in the region; the HK merged catalog contained an extra 16883
resulting in a total of 59675 sources detected in the HK bands alone.
These two catalogs have been used for the photometric analysis
presented in this paper.  The WFCAM photometric system is based on the
system of Mauna Kea Consortium Filters \citep{dye06}.  We have
maintained the same system for plotting the data points, although we
have transformed the data of the reference curves (which are in
Bessel-Brett system) to that of the MKO system.

\subsection{Spitzer Photometry}

Archival Spitzer Space Telescope data were retrieved using the Leopard
software for the analysis presented here. We obtained the Post-Basic
Calibrated Data (PBCD) images from two programs, namely PID-623 (IRAC
Campaign W (SV-3), PI: Giovanni Fazio) and PID-1021 (DR21 and its
Molecular Outflow, PI: Anthony Marston). These programs together cover
almost the entire WFCAM field of view presented here and in Paper I,
encompassing the DR21, W75N, the Diamond ring regions to the south
west of DR21 (see Fig.~1), as well as new sites of star formation to
the west of W75N, and to the west of DR21 (labeled L906E in Fig.~1 and
in Paper I).  Data were obtained with both the MIPS and IRAC cameras,
although here we discuss only data from the latter.  IRAC images in
four filters, known as band 1 (band center wavelength $\sim$3.6\mic ),
band 2 (4.5\mic ), band 3 (5.8\mic ) and band 4 (8.0\mic ) were
obtained.  The IRAC instrument is described by \citet{faz04}.  Both
programs used Full Array Readout with 2~sec exposure times. Mosaics of
the individual frames were made using STARLINK routines and aperture
photometry was extracted using the APPHOT tasks in IRAF. Images and
contour plots in three of the four bands are presented in Paper I.

The PBCD fits files are calibrated in terms of MJy~sr$^{-1}$ and the
photometry tasks used in $iraf$ are designed to handle
counts/sec. Therefore a proper conversion from the flux units to
counts and estimating the photon noise is required.  Photon noise was
computed by using a $phpadu$ (photons per image count), where $phpadu
= (exptime \times gain) / (fluxconv \times pixratio)$.  $exptime$ is
the total exposure time, $gain$ is 3.3, 3.71, 3.8 and 3.8 and
$fluxconv$ is 0.1125, 0.1375, 0.5913 and 0.2008 for channels 1, 2, 3
and 4, respectively. The $pixratio$ is the ratio between the area of
the original IRAC pixels and the pixels in the mosaics. This was 1 in
our case.

Aperture  photometry of point  sources was extracted using an aperture
of 2.4$\arcsec$. Sky correction was applied using an inner sky annulus
of 2.4$\arcsec$ and an outer  sky annulus of 7.3$\arcsec$.  Zero-point
magnitudes corresponding  to  these aperture   settings (including the
aperture corrections) were 17.79, 17.30,  16.71 and 15.89 respectively
for Channels 1, 2, 3  and 4.  Using  the individual photometry we then
obtained matched sources between the   various bands. A total of  1580
point sources  were  matched   between all   four  bands.    The  mean
photometric  uncertainities are 0.02, 0.02, 0.07  and  0.15~mag for
channels 1, 2, 3 and 4, respectively.

\begin{figure}
\vbox to 110mm{\vfil
\includegraphics[width=90mm]{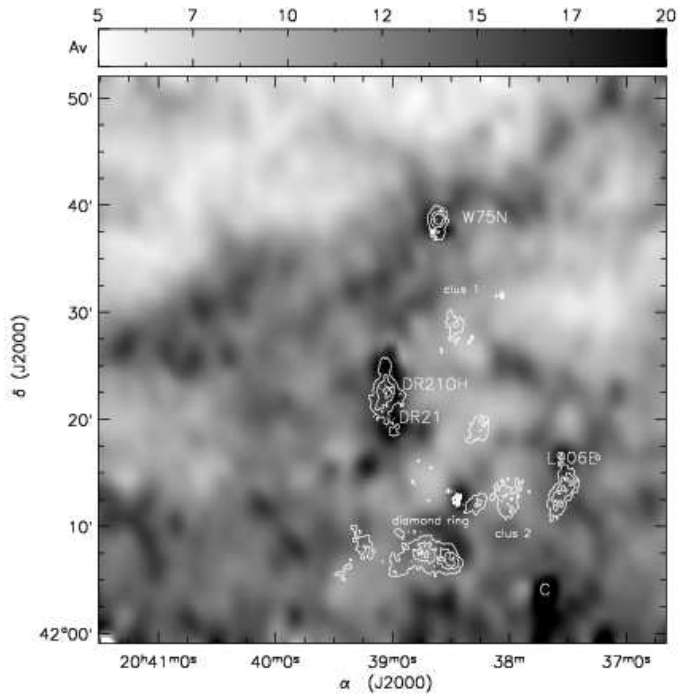}
 \caption{Near-infrared extinction map shown in grey scale. Overlaid
 contours represent nearest neighbour stellar density enhancements as
 derived from the 4.5$\mu$m IRAC image point source photometry
 data. The contour levels show star densities of $\sim$ 30K, 40K, 50K,
 70K, 100K, 150K and 200K stars~degree$^{-2}$ } \label{overlay}
\vfil}
\end{figure}

\begin{figure} \vbox to110mm{\vfil
\includegraphics[width=90mm]{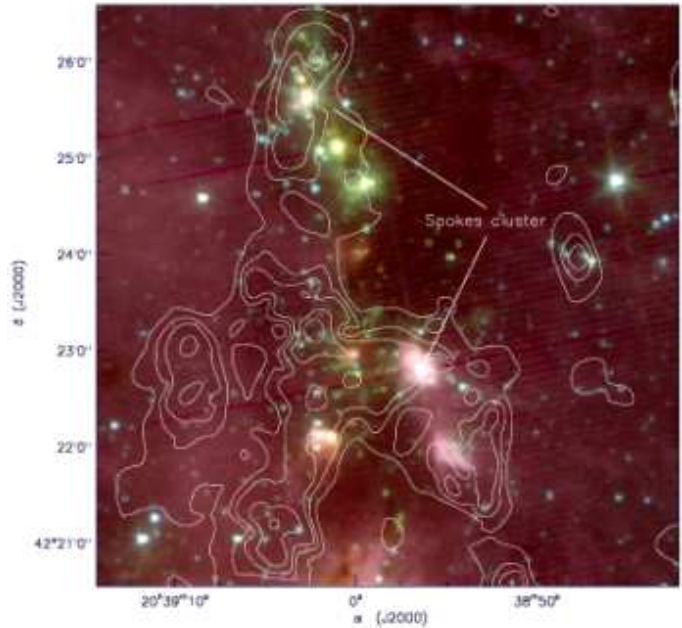} 
\caption{ A zoom-in view of the region around DR21(OH) and the 
associated dense filament, with the three colour composite background
shown in Fig.~1 overlaid by 6th nearest neighbour stellar density
contours.  Note the structure in clustering and the stars that form
spokes cluster. }
\label{clustructure} \vfil}
\end{figure}

\section{Results}

\subsection{Extinction maps}

In Fig.~2, the grey scale background shows the near-infrared
extinction map of the region.  This map was produced by Nyquist
binning the spatial distribution of extinction values with a bin size
of 150$\arcsec$.  The JHK photometric data were plotted on an H-K vs
J-H colour-colour diagram. In this diagram (not shown), all stars were
deredenned to a locus perpendicular to the redenning vector and the
vector length was measured appropriately converted to extinction
values. Given the large distance to this region, the observed area is
crowded by field stars redenned only by interstellar
extinction. Including these field stars affects the contrast of the
extinction map. Therefore, only those points with A$_v\ge$5~mag were
chosen to produce the extinction map shown in Fig.~2. This map traces
regions of extinction between A$_v$ of 5--30~mag, limited by the
completeness limit of the WFCAM observations.  However, the DR21/W75
region are associated with several dense regions with extinction
greater than A$_v\sim$30~mag which do not show any stars at 2\mic ,
resulting in missing data points in our map.  These pockets appear as
white regions adjacent to the black spots (for example see the white
patch adjacent to the black head of the region marked ``C'' in
Fig.~2). Such white patches are supplemented appropriately for the
portion of the field covered by SCUBA observations. This was
accomplished by converting the SCUBA column densities to extinction
values and an absolute calibration made by using robust extinction
data points that was measured with NIR data. In those regions where
SCUBA data is not available the white patches remain. A histogram of
all the extinction values obtained from the WFCAM data suggest that
the mean extinction to the DR21/W75N regions is A$_v$=6-8~mag. This is
in good agreement with the value obtained by
\citet{jos05} towards this line of sight within the galaxy.

\begin{table*}
 \centering
 \begin{minipage}{140mm}
  \caption{ Embedded Clusters in the mapped region }
  \label{embclusters}
  \begin{tabular}{@{}lcccccccc@{}}
  \hline
Region & \multicolumn{2}{c}{Coords (J2000)(deg)} & \multicolumn{2}{c}{ Radius (pc)} &  \multicolumn{2}{c}{Density (stars~pc$^{-2}$)} & Isoperimetric & $\tau$ \\
name & $\alpha$ & $\delta$ & effective & core & half-power & peak & quotient & \\
  \hline
W75N & 309.652 & 42.6462 & 0.96 & 0.63 & 20 & 44 & 0.76 & 0.66 \\
Diamond ring & 309.660 & 42.1197 & 1.62 & 1.62 & 27 & 54 & 0.20 & 0.99 \\
L906E & 309.383 & 42.2238 & 1.24 & 0.91 & 23 & 49 & 0.45 & 0.73 \\
DR21/DR21(OH) & 309.767 & 42.3764 & 1.46 & 1.08 & 24 & 48 & 0.23 & 0.74 \\
\hline
\end{tabular}
\end{minipage}
\end{table*}

The extinction map clearly traces the north-south dense molecular
filament encompassing DR21, DR21-OH and the W75N region (Paper I). It
also reveals other regions of high extinction towards the diamond ring
cluster, L906E and region "C", which is the darkest patch in
Fig.~\ref{overlay} extending beyond the limits of our map.  It can be
noted that these regions of high extinction coincide well with the
brightest emission regions in the MIPS 24$\mu$m image of Marston et
al. (2004) and the 850\mic\ dust continuum cores in Paper I. However,
the Spitzer/IRAC images and 850\mic\ data does not encompass region
``C'', so the associated star formation activity is not discussed
here.

\begin{figure}
\vbox to120mm{\vfil
\includegraphics[width=90mm]{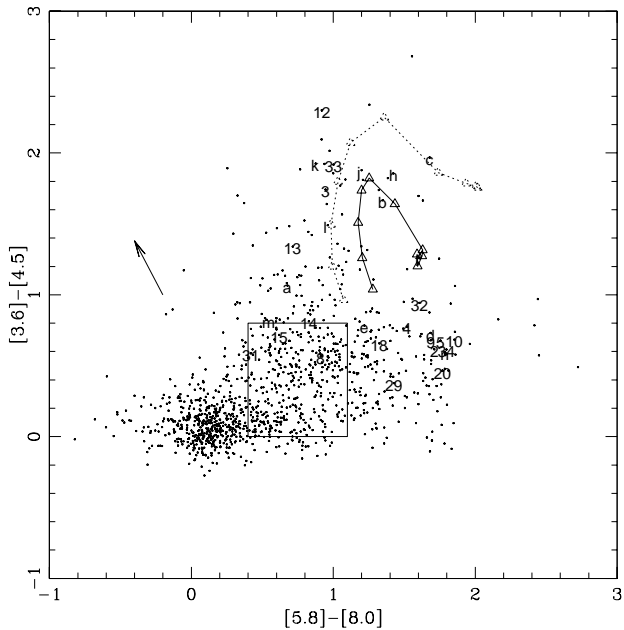}
\caption{ Colour-Colour (CC) diagram from the Spitzer IRAC data for
DR21/W75, including stars associated with the Diamond Ring region and
L906E. The solid box represents the region of Class II sources (Allen
et al 2004). The dotted line shows synthetic fluxes for a YSO with a
temperature of 32000~K; the solid curve shows a YSO track at a
temperature of 15000~K (Whitney et al. 2004).  Numbers and letters
refer to specific sources discussed in Paper I.}
\label{2col}
\vfil}
\end{figure}

\subsection{Embedded Clusters}

The star forming regions DR21 and W75N appear as bright near-infrared
nebulae owing to illuminated gas and shock excited outflows. However,
no significant stellar density enhancements are observed at 2$\mu$m
(in the K band) in these regions.  This means that the embedded
population must be hidden inside the dense molecular cloud at higher
extinction or that the YSO population is too young to be bright at
2$\mu$m.  This supposition becomes obvious with the availability of
IRAC images from the Spitzer Space Telescope, where many embedded
YSO's are clearly visible.  The contours in Fig.\ref{overlay} trace
the embedded clusters in the region as identified by stellar density
enhancements in the 4.5$\mu$m point sources.  These contours were
obtained by computing the 20th nearest neighbour (20NN) enhancements
\citep{casertano85,bruno06} of the spatial distribution of 14,063 sources
detected in this band. The clusters were detected by using the
4$\sigma$ contours above the mean star counts in the region where
$\sigma$ is the variance of the mean star count values. Table.~1 lists
the embedded clusters identified and their properties.  Embedded
clusters associated with DR21/DR21(OH), W75N, the diamond ring region
and L906E were detected. There are a couple of fainter clusters marked
clus1 and clus2 in Fig.~\ref{overlay} but these do not posess
significant peaks, so they are not listed in Table.~1. The embedded
clusters are found to be spatially coincident with the high
extinction regions revealed by the extinction map obtained from the
near-infrared WFCAM data. In fact, the observed clusters are better
aligned with the whitish patches which represent missing data points
in the near-infrared extinction map. The DR21/DR21(OH) cluster also
aligns well towards one edge of the dense filament traced by the
850\mic~emission.

The 4~$\sigma$ contour was used to identify the boundary of the
cluster, its effective radius and isoperimetric quotient. To compute
the peak positions, core and half-power densities and the $\tau$
parameter we made use of a 20NN map. Table.~1 lists effective radius,
peak and average stellar densities, and the $\tau$ parameter for the
clusters. The effective radius is computed by measuring the area $A$
enclosed by the 4~$\sigma$ contour and computing R=$\sqrt{A/\pi}$.  In
Figure.~3 we show contours of a 6NN map to demonstrate the
structure within the clusters; it can be noted that the clusters are
elongated in the direction of the dense filament and they display
significant sub-structure. Two parameters namely the isoperimetric
quotient and $\tau$ are used to measure the structure of the clusters
\citep{bruno06}. The isoperimetric quotient for each cluster is
defined as follows: If the cluster boundary contour has a perimeter
$p$ units enclosing an area $A$ square units, then the isoperimetric
quotient of this curve is $4\pi A/p^{2}$. It measures the ratio of the
area enclosed by the cluster boundary to the area of a circle with the
same perimeter. If our clusters were circular, this quotient will be
equal to 1. $\tau$ is defined as the ratio of the core radius to the
effective radius and measures how much of the cluster population is in
the core and how much is in the halo. \citet{bruno06} defines a
cluster as heirarchical if $\tau$ is greater than 0.455. The last two
columns in Table.~1 lists the isoperimetric quotient and $\tau$ values
for each cluster. It can be noted that the DR21 and diamond ring
clusters are the most elongated and heirarchical, while the W75N
cluster is the most circular and centrally condensed.

\begin{figure}
\vbox to110mm{\vfil
\includegraphics[width=83mm]{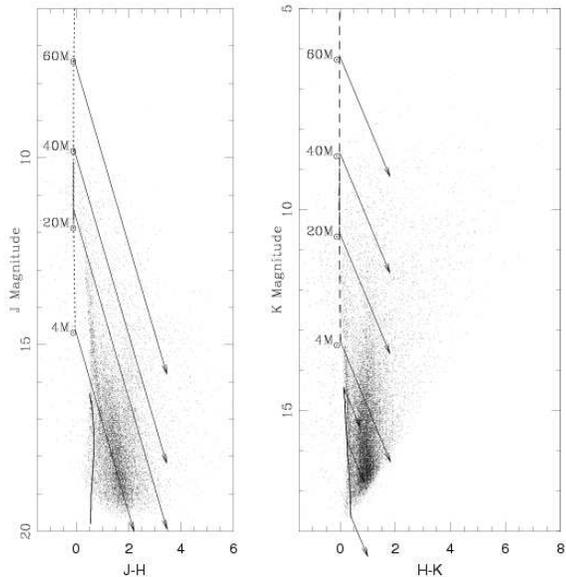}
\caption{Colour magnitude diagrams from near-infrared JHK data. In both 
panels, points represent data derived from WFCAM photometry. The solid
line shows 1Myr  old PMS model  track from Baraffe  et al., the dotted
line represents  the 3Myr  old ZAMS  track from Geneva  models and the
solid arrows represent redenning vectors of A$_v$ = 30~mag.}
\label{starden}
\vfil}
\end{figure}

\subsection{Color-Color(CC) and Color-Magnitude(CM) diagrams}

The nature of the embedded young stellar population can be evaluated
using the photometric data plotted on colour-colour (CC) and
colour-magnitude (CM) diagrams.  The majority of the embedded
population is bright in the IRAC bands.  Therefore, the 1,592 point
sources matched between all four IRAC bands were used to plot the IRAC
CC and CM diagrams.  Fig.~\ref{2col} shows a [5.8-8.0] vs [3.6-4.5] CC
diagram.  The numbers and letters in these plots refer to specific
targets that are discussed in detail in Paper I: the numbers are
bright 8.0\mic\ point sources, often compact clusters and/or luminous
sources in the region, while the letters are candidate outflow driving
sources. Almost all exhibit colours consistent with protostars.  In
Fig.~\ref{2col} the redenning vector is shown with an arrow, the two
curves represent Class I models for two different temperatures
\citep{whitney04}. The dotted curve represents the colours of a
Class I YSO at a temperature of 32000~K for various inclination angles,
and the solid line is a similar object at a temperature of 15000~K.
The square box represents the zone of Class~II objects as discussed by
\citet{all04}. The concentrated circular distribution of points
to the lower-left of the square box represents the pure photospheres.
Fig.~\ref{2col} shows that about 50\% of the point sources detected in
the IRAC bands are pure photospheres, while the remaining are
distributed in the zones representative of younger objects such as
Class 0, I and II \citep{all04,meg04}. Notice the few points with
excess red colours that occupy the upper right-hand regions of
the plot, i.e. regions redder than the model curves of 15~K and 32~K
Class~I source.  These are representative of candidate massive stars.

Although the JHK data do not show any significant embedded clusters,
they do reveal several embedded sources and the majority of the halo
population distributed outside the densest regions of the molecular
cloud.  Also, the higher spatial resolution of the near-infrared data
can separate multiple components (when these are present) among the
embedded population.  As mentioned earlier, the JHK merged catalog
contains $\sim$43,000 sources and $\sim$60,000 sources in the $H$ and
$K$ bands alone.  These points are plotted on two CM diagrams shown in
Fig.~\ref{starden}.  The pre-main-sequence models of \citet{baraffe98}
(solid line) and the zero-age-main-sequence model tracks from the {\em
Geneva Observatory} stellar models (dotted line)\citep{lejeune01} are
plotted for reference. The observed points were compared to model data
with ages between 1~Myr--3~Myr and distances ranging between 2--3~kpc.
We find that the near-infrared data are matched relatively well for an
age of 3~Myr and a distance of 3~kpc.  Note that the age of 3~Myr
applies only to the population of stars detected in the NIR JHK
bands; it does not apply to the deeply embedded population revealed by
the IRAC data. The YSO population traced by IRAC data consists largely
of younger objects, with an age of perhaps 1~Myr or less.

\begin{figure} 
\vbox to110mm{\vfil
\includegraphics[width=90mm]{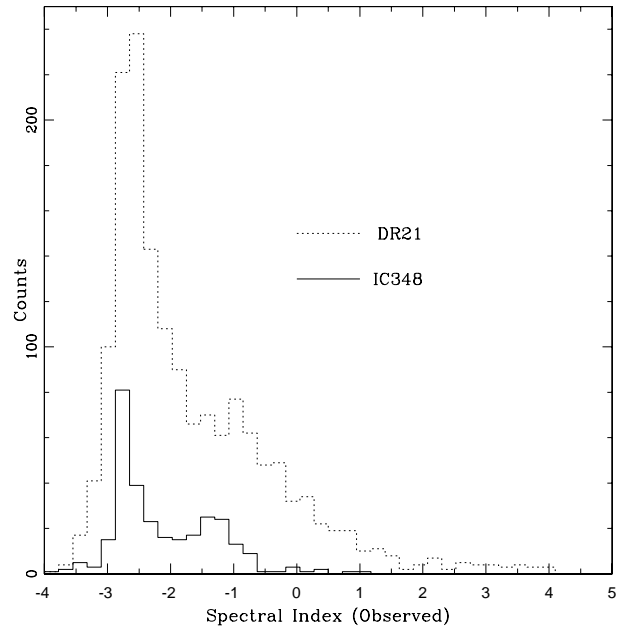} 
\caption{ A histogram of spectral indices measured from linear
least-squared fits to source photometry in the four IRAC bands.
Thus, only sources detected in all four bands are plotted.  A similar
plot is presented for the IC348 Spitzer observations of Lada et
al. (2006) for comparision.}
\label{absmagcol1} \vfil}
\end{figure}

\subsection{Spectral indicies of the embedded population}

The IRAC bands are well suited to measuring the infrared excess
emission that can be attributed to circumstellar envelopes and disks.
The IRAC band fluxes have been successfully used to determine the disk
excess objects and obtain reliable estimates of the disk fraction in
nearby low mass star forming regions such as IC348 \citep{lada06}
. \citet{lada06} show that the pure photospheres and the disk excess
sources clearly separate out when the IRAC fluxes are used to estimate
the power law index of the observed spectral energy distribution.
This is because the disk emission dominates in the 4.5-8.0$\mu$m,
where the SED's will significantly deviate from the pure photospheres
of low mass stars. A simple least squares linear fit to the observed
SED between 3.6$\mu$m and 8.0$\mu$m was obtained for all of the 1,580
sources found in all four IRAC bands.  Table.~2 (available online)
lists the magnitudes in the four bands and the computed spectral
indices.  The SED's were not deredenned before fitting because an
accurate estimate of the A$_v$ to each source is possible only by
fitting model atmospheres. This can not be done with the limited data
available for this region.

A histogram of the observed spectral indicies $\alpha$ for all sources
detected in the four IRAC bands in DR21/W75 is shown in
Fig.~\ref{absmagcol1}.  For comparision we also plot the indices
measured for IRAC sources observed in the low mass star forming
region, IC348, as described by \citet{lada06}. DR21/W75N is located at
3\,kpc resulting in a large line-of-sight extinction that will stretch
the $\alpha$ values. Extinction due to a fixed column density of gas
and dust is a function of wavelength which will attenuate the shorter
wavelengths much more than the longer wavelengths. This will result in
an artificial rise in the slope of the spectral energy distribution if
the foreground extinction is significant, resulting in a higher value
of $\alpha$. However, the change is larger for sources with lower
$\alpha$ and relatively smaller for sources with higher $\alpha$ thus
causing a stretching of the $\alpha$ values rather than a simple
increase. We therefore corrected the IRAC magnitudes for such effects
by using the standard extinction law \citep{mathis90} and assuming an
uniform foreground extinction value of Av=9\,mag. Note, however, that
it does not correct the actual extinction to each source which is due
to the molecular cores or disks. In DR21/W75, the observed spectral
indicies range from -4.0 to +4.3. The histogram for IC348 shows a
prominent peak at $\alpha\sim$-2.7, which corresponds to pure
photospheres, and a somewhat less prominent peak at $\alpha\sim$-1.8,
corresponding to sources with thick disks \citep{lada06}. Similar
peaks can be seen for the DR21 data as well, the shapes of both
histograms compare quite well, representing similar aspects of star
formation in both regions.  However, the histogram for DR21 deviates
from that of IC348 at $\alpha$ values greater than 0 and particularly
beyond 2. A small, but significant fraction of sources in DR21 are
found to have high spectral indicies, between 2 and 4.5. The higher
values of $\alpha$ is not merely an effect of the stretching effect
explained above due to high density regions. The high $\alpha$ sources
are better represent the source functions because the change
$\delta\alpha$ is smaller for high $\alpha$ sources compared to low
$\alpha$ sources. We shall discuss the implications of this result in
the next section.

\begin{figure*}
\vbox to250mm{\vfil
\includegraphics[height=220mm]{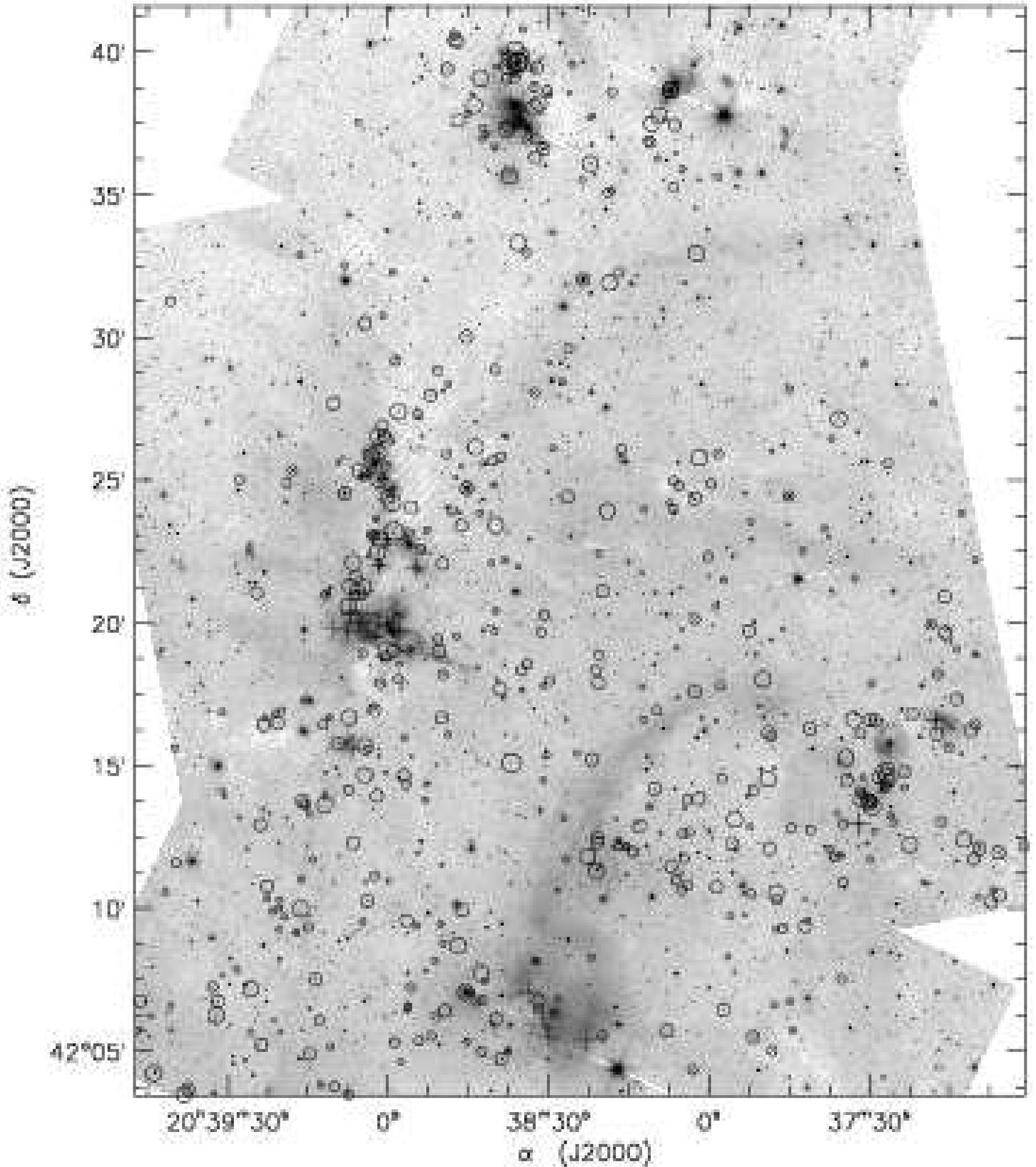}
\caption{Distribution of infrared excess sources. IRAC channel 2 image 
is  shown in grey scale and  the circular symbols represent the excess
sources. Crosses represent sources with $\alpha >$3. The size of the symbols are proportional to  the value of the
spectral index.}
\label{irac} \vfil}
\end{figure*}

In Fig.~\ref{irac} we show the spatial distribution of the IRAC point
sources using circular symbols. Sources with $\alpha >$3 are shown by
cross marks. The size of each symbol is proportional to the observed
spectral index of the corresponding source. The background greyscale
image is a 4.5$\mu$m mosaic image of the region.  Sources with the
highest spectral index are concentrated in the densest regions
associated with the north-south molecular ridge that runs through DR21
(see Fig.~1), W75N, L906E in the west-south-west, and the Diamond Ring
region. In particular, strong clustering of sources can be noticed
around DR21, DR21OH, W75N and L906E region (see Fig.~\ref{jhk} and
Fig.~7). These clusters are connected by a distributed population of
sources with high spectral indices, that can be found spread along two
filaments, one running north-south and the other towards the
south-east and north-west. A general correlation of the disk sources
with the dark patches in the entire field can be seen in
Fig.~\ref{irac}.

\section{Discussion}


We have explored the stellar population in and around two major,
luminous \citep[$\sim10^5$\Lsolar ][]{moo88,cam82} star forming
regions, namely DR21 and W75N, using wide field UKIRT/WFCAM JHK and
Spitzer/IRAC 3.6, 4.5, 5.8 and 8.0$\mu$m images.  The observed area
encompasses several luminous far-infrared sources, clusters of UCHII
regions \citep{tor97,cyg03}, sites of intense maser activity
\citep{pla90,man92,hun94,kog98,tor97} and powerful molecular outflows
\citep{gar91,dav98}, all of which are signposts of
massive star formation.  The estimated distance to all these sources
is more or less the same; the analysis of the data presented in
Sec.~3.3 suggests a distance of $\sim$3~kpc, which is in good
agreement with other estimates derived from various methods
\citep{gen77,cam82,fis85,ode93}.


The results presented in the previous section demonstrate that the
DR21, DR21(OH) and W75N regions contain young stellar objects that can
be classified as Class 0/I or Class II, based on Spitzer colours and
spectral indices. This implies that star formation is a recent
phenomenon in all of these regions.  The majority of the young stellar
population in each area is so deeply embedded that it is not uncovered
in observations at wavelengths shortward of 2$\mu$m. The SCUBA
850\mic\ cores described in Paper I and shown here in Fig.~1 probably
represent an evolutionary phase that is earlier even than that traced
here with Spitzer.  The SCUBA cores (32 are identified in Paper I) are
typically not associated with 8.0\mic\ sources.  They are generally
confined to a narrow, north-south filament in DR21, and to a compact
region around W75N.  The Spitzer YSOs, though also certainly clustered
around the dense cores and filaments, are none-the-less somewhat more
widely distributed.  This could be due to the displacement of YSOs
away from the central density peaks as they evolve.  Even so, the
different epochs of star formation traced by WFCAM, Spitzer and SCUBA
all seem to be co-existing in DR21/W75. Moreover, common trends such
as distance, YSO class and estimated age suggest that the individual
sites of star formation -- W75N, DR21, DR21(OH), the Diamond Ring and
L906E -- may be connected in a global way.  In the following we shall
argue that there is more evidence to support this scenario.


The 8$\mu$m channel of the IRAC instrument traces warm dust and
emission from PAHs (polycyclic aromatic hydrocarbons) much more
efficiently than the shorter wavelength bands. The colour image of the
entire region shown in Fig.~1 displays prominent filamentary nebulae.
At least five long filaments of nebulosity (marked by the sketched
lines in Fig.~1) can be identified in the observed region, four of
which converge on the DR21/DR21OH regions. Note that the bases of
these filaments are traced to some extent even in the 850\mic\
emission, showing that the filaments are indeed made of dense gas. We
have seen from Fig.~\ref{irac} that a visibly prominent distribution
of thick disk sources is concentrated along the thickest filament
coinciding with the DR21, DR21OH and FIR~1/2/3 regions, although the
W75N region is well away from these filaments and appears
disconnected. A careful examination of the same figure will also show
that two other filaments that originate at DR21 and run towards the
northwest and a third directly westward of DR21 are also correlated
with the distribution of thick disk sources. The thin filament running
directly westward of DR21 and linked to the outflow may be due to the
outflow punching out of the filament. While the correlation of
embedded YSOs and filaments is prominent for the filaments converging
on DR21, the same is not true for the diamond ring region. The diamond
ring region has a bright filament running northwards that curves
toward the L906E region marked in Fig.~1. The YSOs in Fig.~7 all
appear to be concentrated in the L906E region and just below the
filament, rather than at the head of the diamond ring. This suggests
that the diamond ring region is relatively evolved, while L906E is
actively forming young stars.  Note that no H$_2$ jets were discovered
near the diamond ring, while a number of flows were identified around
L906E (Paper I).  A proper analysis of the filaments and their
relative orientation requires additional mapping at 850\mic\ (our
SCUBA map does not extend as far south as the diamond ring, or as far
west as L906E) and CO line data with good spatial and velocity
resolution.

As shown in section.~3.2, embedded clusters are associated with each
of the main targets in the region, namely DR21/DR21(OH), W75N, L906E
and the diamond ring region. The clusters show cores on scales of
$\sim$1~pc. If each cluster peak represents a molecular clump out of
which the clusters were born, then the size scales of the cluster
cores represent the scales of fragmentation in this filamentary
cloud. The main DR21 filament traced by the 850\mic\ emission is about
16~pc long and is one of the most structured filaments in terms of
embedded clusters. A zoom-in view of this region is shown in Fig.~3
along with contours of the stellar density. The contours display
multiple stellar density peaks.  These clusters are elongated roughly
along the north-south filament.  In this figure notice the bright
stars marked as ``Spokes Clusters'', which include the FIR 1/2/3
850\mic\ peaks (refer Fig.~1) in one line. Each of these bright stars
are surrounded by a noticeable linear alignment of fainter stars that
are arranged like the spokes of a wheel whose centre coincides with
the bright star.  This type of structure has been recently identified
by \citet{paula06} in NGC2264, which is at a somewhat closer distance
of d=800~pc, and is therefore more clearly visible. These authors show that
the linearly aligned fainter stars are indeed redder than the central
luminous source. Such structures are thought to be important because
they represent primordial structures in the formation of a star
cluster representing true proto-clusters
\citep{kurosawa04,bate03}. Although the spokes configuration in the
clusters shown in Fig.~3 are not as prominent as that shown by
\citet{paula06} (possibly due to the larger distance of DR21), there
are never-the-less four such clusters in DR21. The association of
these clusters with bright 850\mic\ emission and a young population of
stars with high spectral indicies indicates that this region may
represent multiple adjacently located protoclusters.

The filamentary structures found in the DR21/W75 region on the scales
of giant molecular clouds (16-20~pc), and those found on smaller
scales, such as the ``spokes clusters'', collectively resemble the
pattern predicted by the numerical simulations of \citet{bate03}. The
simulations by \citet{bate03} are made for a 50\msun\ cloud, which is
much smaller than the actual mass of the DR21/W75 clouds. However,
qualitatively, their results seem to be applicable even here,
suggesting that the mechanism of gravo-turbulent star formation
effectively operates on the scales of giant molecular clouds, up to
several hundred thousand solar masses and several parsecs in
dimension. The filamentary nature appears heirarchical if we consider
the smaller scale spokes configurations and the larger scale filaments
feeding on to the center of the DR21/DR21(OH) region. This heirarchy
can be carefully observed by comparing structures in Fig.~1 and
Fig.~3.

Finally, we mention the implications of the sources with high $\alpha$
values in DR21 distributed along the dense filament and also
coinciding with sign-posts of massive star formation. $\alpha >$2 in a
simplistic interpretation implies presence of Class 0 objects.
Recently, \citet{robitaille06} presented and discussed a large grid of
200,000 YSO model SEDs. According to their work, there are a number of
reasons why $\alpha$ (measured in the Spitzer IRAC bands) can assume
high values: a) younger sources such as Class 0 sources are known to
have high $\alpha$, b) higher mass accretion rate and/or larger
disk/envelope masses boost $\alpha$, and c) beyond 5000K, $\alpha$
will increase with the temperature of the star, due to a lower
contribution from the stellar flux and a higher contribution from the
infrared dust spectrum.  In DR21/W75N the observed data points satisfy
almost all of the above conditions because they are situated in the
densest parts of the star forming region, they coincide with
sign-posts of massive star formation such as UCHII region and water
masers, and because they occupy the colour zone of Class 0 protostars
on a CC diagram. More quantitative comparisions of the real nature of
massive stars requires the plotting of data from theoretical SEDs on to CM
diagrams. However, in colour-colour space there is significant overlap
of data points for the different evolutionary states, source masses,
temperatures, disks and even inclination angles
\citep{robitaille06}, particularly if one uses a limited number of
photometric bands.  Rigorous comparisions will require multiple
wavelength observations, particularly at larger wavelengths such as
24\mic\ and 100\mic\ where younger and luminous objects are known to
particularly bright.

\section{Conclusions}

We analyse NIR and MIR point-source photometry over an area that
includes the massive star forming regions DR21, DR21(OH), and W75N, as
well as recently-discovered clusters of more evolved young stars (the
Diamond Ring region), and bok-globule-like cores where low mass stars
are probably forming (L906E). We use colour-colour and
colour-magnitude diagrams, as well as extinction maps and surface
density plots, to examine the population of young stars across the
region.  We also measure the MIR spectral index for 1580 sources.

Our extinction map traces much of the clumpy, north-south filament
associated with DR21 and DR21(OH), as well as new cores that may be
associated with separate episodes of star formation.  The very densest
cores, however, are not extracted from the NIR data, because of
extinction larger than A$_v \sim$30~mag.

NIR and MIR color-colour and colour-magnitude plots show that the
JHK and 4.5\mic --8.0\mic\ data trace different YSO populations, as
expected: the WFCAM data reveal sources with ages of $\sim$1--3~Myr,
and mostly present in a widespread halo, while the IRAC point sources
are clustered along the dense filaments traced by 850\mic\
data. Spectral indices for the Spitzer sources indicate a more
embedded and/or massive population in DR21/W75 than is seen in the
IC348 low-mass star forming region.  There is a significant fraction
of sources with indices greater than 2 and up to 4 in DR21/W75.
Indeed, $\sim$40\% of the Spitzer-detected sources in the field have
SEDs consistent with Class 0/I or Class II YSOs. The reddest sources
are generally associated with the densest cores, as traced in 850\mic\
dust continuum emission in Paper I, or in 8.0\mic\ PAH emission. The
NIR data, tracing the evolved and halo YSOs represent a $\sim$\,3\,Myr
old population whereas the Spitzer data, tracing the more deeply
embedded content and also associated with outflow activity, represents
a relatively younger population of 1\,Myr or less.
 
Star formation in DR21 region is found to be occuring in filaments of
10-20\,pc lengths that is fragmented in to multiple clusters on the
scales of 1-2\,pc. We identify four spokes cluster configurations in
the main DR21 filament that is representative of proto-clusters. The
morphological appearance on the scales of the entire Giant Molecular
Cloud resembles well the numerical simulations of cluster formation by
\citet{bate03}.

\section{Acknowledgments}

MSNK greatfully acknowledges Paula Teixeira for helpful discussions on
Spitzer photometry.  We also thank an anonymous referee for useful
suggestions and a speedy review. MSNK and JMCG are supported by a
grant POCTI/CFE-AST/55691/2004 and JMCG is supported by a grant
SFRH/BD/21624/2005 approved by FCT and POCTI, with funds from the
European community programme FEDER.  DF received funding from the
CosmoGrid project, funded by the Program for Research in Third Level
Institutions under the National Development Plan and with assistance
from the European Regional Development Fund.  This research made use
of data products from the Spitzer Space Telescope Archive. These data
products are provided by the services of the Infrared Science Archive
operated by the Infrared Processing and Analysis Center/California
Institute of Technology, funded by the National Aeronautics and Space
Administration and the National Science Foundation.  Finally, we would
like to thank the team at CASU for processing the NIR data, and the
WFCAM Science Archive in Edinburgh for making the data available to
us.

\label{lastpage}

\end{document}